\def \SAIT #1 #2 {{\em Mem.\ Soc.\ Astron.\ It.\/} {\bf #1}, #2}
\def \MESS #1 #2 {{\em The Messenger\/} {\bf #1}, #2}
\def \ASTRNACH #1 #2 {{\em Astron. Nach.\/} {\bf #1}, #2}
\def \AAP #1 #2 {{\em Astron. Astrophys.\/} {\bf #1}, #2}
\def \AAL #1 #2 {{\em Astron. Astrophys. Lett.\/} {\bf #1}, L#2}
\def \AAR #1 #2 {{\em Astron. Astrophys. Rev.\/} {\bf #1}, #2}
\def \AAS #1 #2 {{\em Astron. Astrophys. Suppl. Ser.\/} {\bf #1},
#2}
\def \AJ #1 #2 {{\em Astron. J.\/} {\bf #1}, #2}
\def \ANNREV #1 #2 {{\em Ann. Rev. Astron. Astrophys.\/} {\bf
#1}, #2}
\def \APJ #1 #2 {{\em Astrophys. J.\/} {\bf #1}, #2}
\def \APJL #1 #2 {{\em Astrophys. J. Lett.\/} {\bf #1}, L#2}
\def \APJS #1 #2 {{\em Astrophys. J. Suppl.\/} {\bf #1}, #2}
\def \APSS #1 #2 {{\em Astrophys. Space Sci.\/} {\bf #1}, #2}
\def \ASR #1 #2 {{\em Adv. Space Res.\/} {\bf #1}, #2}
\def \BAIC #1 #2 {{\em Bull. Astron. Inst. Czechosl.\/} {\bf #1},
#2}
\def \JSQRT #1 #2 {{\em J. Quant. Spectrosc. Radiat. Transfer\/}
{\bf #1}, #2}
\def \MN #1 #2 {{\em Mon. Not. R. Astr. Soc.\/} {\bf #1}, #2}
\def \MEM #1 #2 {{\em Mem. R. Astr. Soc.\/} {\bf #1}, #2}
\def \PLR #1 #2 {{\em Phys. Lett. Rev.\/} {\bf #1}, #2}
\def \PASJ #1 #2 {{\em Publ. Astron. Soc. Japan\/} {\bf #1}, #2}
\def \PASP #1 #2 {{\em Publ. Astr. Soc. Pacific\/} {\bf #1}, #2}
\def \NAT #1 #2 {{\em Nature\/} {\bf #1}, #2}

\documentstyle{memsait}
\input epsf.sty
\begin{opening}
\title{Lithium and binarity in M67}
\author{D. Barrado y Navascu\'es$^1$, J.R. Stauffer$^1$, L.
Hartmann$^1$, S. Balachandran$^2$}
\institute{$^1$Center for Astrophysics, Cambridge, USA\\
$^2$ University of Maryland, College Park, USA}
\date{} 
\end{opening}

\begin{document}

\oddpagefooter{}{}{} 
\evenpagefooter{}{}{} 
\ 
\bigskip

\begin{abstract}

A study of the lithium abundances in binaries of the old open
cluster M67 is presented. Abundances were estimated using curves
 of growth and equivalent widths of the LiI6707.8 \AA{ } doublet.
We have  corrected of the effects of the companion on the
measured  equivalent width and on the color indices by deconvolving 
the photometry, computing colors and magnitudes for each
component.

As happens in the Hyades cluster, there are some lithium excesses
in main--sequence and
evolved  TLBS binaries with  respect other binaries and single
stars, although the M67 data have large errors.

\end{abstract}

\section{ The data}

We present a study of lithium abundances in M67 binaries,
selected from 2
different  photometric studies: Sanders (1989) and Montgomery et
al. (1993).
Our sample covers systems in a variety of evolutionary stages
and orbital periods.

Three different observing runs were carried out at Kitt Peak
National 
Observatory  (Jan 1994, Jan 1996 and Jan 1997), using the Mayall
4m telescope. We also used the 2.5m INT and the
4.2m WHT telescopes (March 1996 in the first case, May 1996  and
March 1997 in the second) at the "Roque de los Muchachos"
Observatory.
The final resolutions were R$\sim45\,000$, $50\,000$ and
$12\,000$, for the KPNO, WHT and INT data.

Binaries are, on average, brighter that single stars and/or than
the cluster isochrone at the same (B--V) value.
Following Barrado y Navascu\'es \& Stauffer (1996), we have
deconvolved the photometry of our binaries to obtain colors,
magnitudes of their components and continuum correction factors. 
The actual deconvolution process consists on the calculation of
fictitious binaries using the photometric  data from the
isochrone, adding pairs of stars as primary and secondary. The
comparison between the calculated  and the observed data provides
the best solution for the deconvolution.

\begin{figure}
\vspace{5cm}
\includegraphics{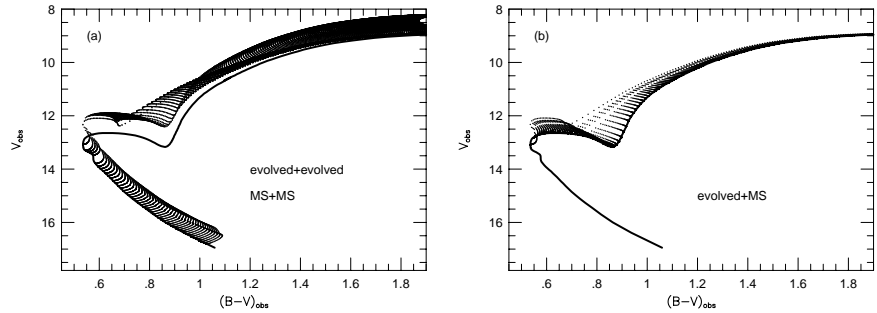}
\caption[h]{M67 isochrone in the Color-Magnitude Diagram
 (wide solid line) and the location of computed binaries. 
{\bf a.-} Binaries formed by 
MS components only or by evolved stars only. 
{\bf b.-} Systems having an
evolved primary and a MS secondary.}
\end{figure}

Fig. 1 shows the computed location of binaries in the V-(B--V)
plane. The wide solid line represents a  4$\times$10$^9$ yr old
isochrone, using  (m--M)=9.60, E(B-V)=0.04,
and the dotted lines  corresponds to different series of
calculations.  Although the
deconvolution of the photometry can be very powerful, our results
should be taken with some caveats due to the uncertainties in the 
original data.

In order to estimate the lithium abundances of our sample of 
stars, we measured equivalent widths of the Li{\sc I}6707.8 \AA{
} feature.  Since some of these  binaries have poor or
unknown ephemerides, we measured the wavelength of very prominent
lines in the spectral range around the lithium feature  for the
primary component in the case of SB1
spectra, and for both, in the case of the SB2 systems. Then, 
we identified the Fe{\sc I}6707.4 \AA{ } and Li{\sc I}6707.8 
\AA{ } lines. In some cases, due to the resolution and/or the
Doppler broadening,  both features were blended. However, we
normally  did not have problems to deblend them by fitting
Gaussian curves  and to obtain the respective equivalent widths.
When this process was not possible, we only measured the total
equivalent  width and eliminated the contribution of the FeI line
by measuring the equivalent widths of other iron lines.  Then we
used empirical relations to determine the
contribution.
 We estimated the effective temperatures using  Thorburn et al.
(1993)
scale.

Preliminary lithium abundances were estimated using curves of
growth. We selected the Pallavicini et al. (1987) set, adding a
curve from 
Soderblom et al. (1993) for T$_{\rm eff}$=4000 K.
 These abundances are still tentative, since we are
improving the corrections  of the spectral continuum and the
colors due to the presence of a companion.

\begin{figure}
\vspace{5cm}
\includegraphics{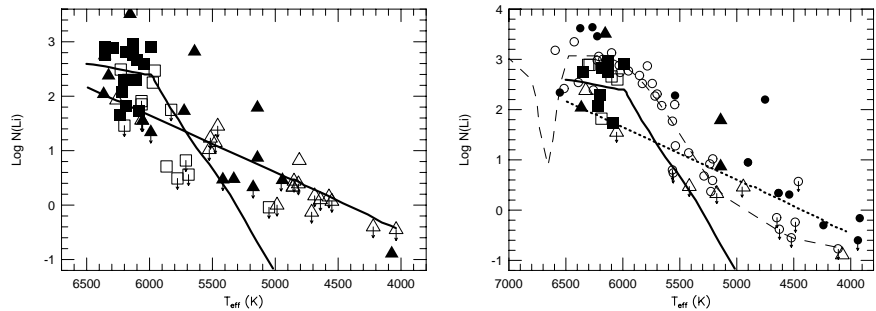}
\caption[h]{ Li abundances against effective temperatures.
{\bf a.-} Single stars are shown as open symbols and  binaries as
solid ones (triangles for evolved systems and squares MS).
{\bf b.-}  Hyades (circles) and M67 binaries (triangles). 
Solid symbols represent synchronized 
binaries, whereas open ones show other binaries 
The solid, dashed and dotted lines represent the behavior of
main--sequence M67, evolved
M67 and main--sequence Hyades single stars. }
\end{figure}

\section{ Lithium  in the Main Sequence and beyond} 

Fig 2a shows   the T$_{\rm eff}$--Log N(Li) plane. 
 Some evolved binaries
have higher abundances than similar single stars.
 Almost half of the sample of main--sequence
 binaries have abundances close to their 
initial level --Log N(Li)$\sim$3.2--. This phenomenon is clearly
 illustrated in Fig. 3, which contains
the normalized histograms for the lithium abundances of 
main--sequence binary and 
single  stars. We selected stars having temperatures higher than
5900 K.  Although the sample is not large, 
both the minimum and the maximum abundances of binaries
are larger than their counterparts for the single stars, and the
values of the abundances of binaries show a peak around Log
N(Li)$\sim$2.9.  Tidally Locked Binary Systems (TLBS) belonging
to the Hyades   have   larger abundances than its singles members
(Barrado y Navascu\'es \& Stauffer 1996),  and
 binaries with no coupling between the
orbital and the rotational periods have over and under abundances
when comparing with single stars.
Fig. 2b includes  data corresponding to binaries from the Hyades
 and M67. For Hyades binaries, we have assumed that those systems
having P$_{\rm orb}\le$9 days are synchronized, since 
they are expected to arrive  the Zero Age Main Sequence (ZAMS)
with equal rotational  and  orbital periods. In the case of M67
binaries, we have
assumed synchronization for MS binaries if  P$_{\rm orb}\le$10
days, whereas  evolved M67 binaries
with P$_{\rm orb}\le$20 days are in fact TLBS. This happens because
they can synchronize the
rotation with the orbital period during the main--sequence 
life--time or during the evolution off the MS, when deep
convective external layers are developed.

Fast rotators belonging to the Pleiades  (70--120 Myr)
have larger
 abundances than slow ones  (Garc\'{\i}a--L\'opez et al. 1994), 
but  the few Pleiades TLBS have 
abundances compatible with their single counterparts. This fact
probably rules out that  the lithium
overabundance in TLBS develops during the PMS life--time.

Hyades TLBS appear to show  Li excesses for objects in the range 
6500--4000 K, when comparing with longer period binaries or
single stars.  In the case of M67 dwarf binaries, they appear to
have Li excesses. However, several TLBS have  abundances below the
{\it maximum} values of single stars. There are
also several binaries  having longer orbital periods and
abundances  close to the maximum values of the TLBS, demonstrating
that there is no clear link between Li and P$_{\rm orb}$ for old binaries.
In any case, the average abundance is larger in the case of TLBS
than other binaries or single  stars.

A preliminary conclusion can be drawn from Fig. 2b,
that Li depletion is inhibited in M67 TLBS, although
the reason why is not clear.  On one hand, these possible
excesses could appear during the first gigayear (till the Hyades'
age or a little longer) due to the inhibition related to rotation
(see discussion in 
Barrado y Navascu\'es \& Stauffer 1996) and they could decrease
afterwards. On the other hard, these M67 stars are quite close to
the turn off point and to the Li dip. The fast and important
evolution of the internal structure which take place at that
moment, together the uncertainties in T$_{\rm eff}$, make the
 interpretation of this phenomenology difficult.

\begin{figure}
\vspace{5cm}
\includegraphics{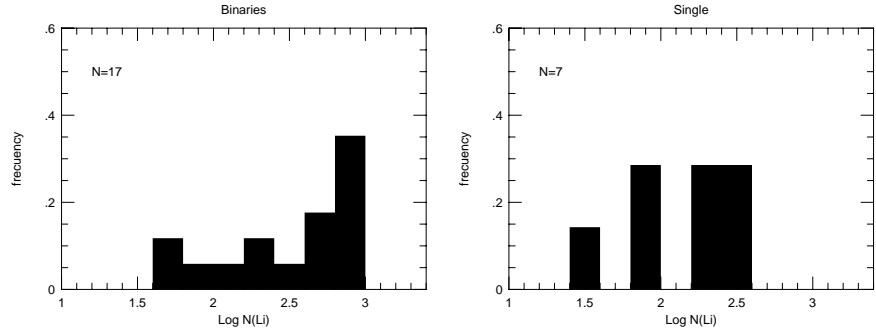}
\caption[h]{Histograms of the distribution of lithium abundances
in main--sequence M67 stars (6500 $>$ T$_{\rm eff}$ $>$ 5900 K). 
{\bf a} Binaries. {\bf b} Single stars.}

\end{figure}

Fig. 2b shows that binarity could be an important factor in the
evolution of lithium in evolved stars.  All four evolved TLBS
have been synchronized on the MS  (or, perhaps, during the
evolution  along the giant gap).
They show Li in their spectra, and their abundances are larger
than the average values for their temperatures.

This  fact could
be interpreted as a prove that quasi-synchronous  systems  at ZAMS
(binaries
close to synchronization, but with P$_{\rm orb}$ slightly 
larger than 8 days) have enough transfer of angular momentum to 
prevent some lithium depletion.

\section{conclusions}

This study  establishes several preliminary conclusions:
 MS binaries  seem to have larger abundances, on average, than
their single counterparts. These excesses should be tested by
more exhaustive analysis.  
 These excesses could appear between Pleiades
and Hyades' age.
 We have  also  found lithium in several TLBS giant binaries.

\acknowledgements
DBN thanks the Real Colegio
Complutense at Harvard University and the MEC/Fulbright program.
 JRS acknowledges support from NASA Grant NAGW-2698.

\end{document}